\documentclass[twocolumn,superscriptaddress, nofootinbib]{revtex4-1}
\usepackage{amsmath}
\usepackage{amsfonts}
\usepackage{amssymb}
\usepackage{float}
\usepackage{graphicx}
\usepackage{calc}
\usepackage{color}
\usepackage[normalem]{ulem}
\usepackage{enumitem}

\newcommand{\sub}[1]{\ensuremath{_{\textrm{#1}}}} 
\newcommand{\super}[1]{\ensuremath{^{\textrm{#1}}}} 

\begin{document}
\title{Hydrodynamic and Ballistic AC transport in 2D Fermi Liquids}
\author{Mani Chandra}
\email{mani@quazartech.com}
\affiliation{Research Division, Quazar Technologies, Sarvapriya Vihar, New Delhi, India, 110016}

\author{Gitansh Kataria}
\affiliation{Research Division, Quazar Technologies, Sarvapriya Vihar, New Delhi, India, 110016}

\author{Deshdeep Sahdev}
\affiliation{Research Division, Quazar Technologies, Sarvapriya Vihar, New Delhi, India, 110016}

\author{Ravishankar Sundaraman}
\affiliation{Department of Materials Science and Engineering, Rensselaer Polytechnic Institute, Troy, NY 12180}

\begin{abstract}

Electron transport in clean 2D systems with weak electron-phonon (e-ph) coupling can transition from an Ohmic to a ballistic or a hydrodynamic regime. The ballistic regime occurs when electron-electron (e-e) scattering is weak whereas the hydrodynamic regime arises when this scattering is strong. Despite this difference, we find that vortices and a negative nonlocal resistance believed to be quintessentially hydrodynamic are equally characteristic of the ballistic regime. These non-Ohmic regimes cannot be distinguished in DC transport without changing experimental conditions. Further, as our kinetic calculations show, the hydrodynamic regime in DC transport is highly fragile and is wiped out by even sparse disorder and e-ph scattering. We show that microwave-frequency AC sources by contrast readily excite hydrodynamic modes with current vortices that are robust to disorder and e-ph scattering. Indeed, current reversals in the non-Ohmic regimes occur via repeated vortex generation and mergers through reconnections, as in classical 2D fluids. Crucially, AC sources give rise to strong correlations across the entire device that unambiguously distinguish all regimes. These correlations in the form of nonlocal current-voltage and voltage-voltage phases directly check for the presence of a nonlocal current-voltage relation signifying the onset of non-Ohmic behavior as well as also for the dominance of bulk interactions, needed to confirm the presence of a hydrodynamic regime. We use these probes to demarcate all regimes in an experimentally realizable graphene device and find that the ballistic regime has a much larger extent in parameter space than the hydrodynamic regime.
\end{abstract}

\maketitle

\section{Introduction}

Charge transport in conductors is typically dominated by electrons scattering against phonons and defects, resulting in momentum relaxation (MR) over a time scale $\tau\sub{mr} \sim 10^{-14}-10^{-15}$~s and a corresponding length scale $l\sub{mr} \sim 10^{-2}-10^{-3}$ $\mu$m. In comparison, the time scale $\tau\sub{mc} \sim 10^{-12}$~s of momentum-conserving (MC) scattering (primarily due to electron-electron (e-e) scattering\footnote{Note that electron-electron Umklapp scattering relaxes momentum, while small-angle electron-phonon scattering at low temperatures does not; we work with momentum-conserving and momentum-relaxing time-scales to avoid this ambiguity.}) is negligible \cite{AshMer}, with the corresponding length scale $l\sub{mc}\sim 1$ $\mu$m. Electrons scattering against phonons or defects give rise to the characteristically diffusive Ohmic transport. However, in clean systems with weak electron-phonon scattering, the length scale of momentum relaxation can approach the device scale\footnote{We are considering devices with dimensions slightly smaller than $l\sub{mr}$ but much larger than the phase coherence length scale so as to preclude interference effects.} (e.g., $l\sub{mr}\sim$ a few $\mu$m in graphene \cite{GrapheneMicronBallistic}). Electrons in this \emph{ballistic} regime scatter predominantly against device boundaries. If e-e scattering is then increased upon adjustment of experimental conditions such as carrier concentration and/or temperature such that $l\sub{mc}$ is made sufficiently smaller than the device scale, electron transport can be \emph{hydrodynamic}. In this regime of slow MR \emph{and} fast MC scattering\footnote{Our discussion pertains to Fermi liquids. Hydrodynamics is then a regime of weakly interacting (so that quasiparticle excitations are well-defined), but fast MC interactions.}, electrons are expected to move collectively as in a fluid whose dynamics are governed by macroscopic conservation laws, i.e., the Navier-Stokes equations. Candidate materials for hydrodynamic charge transport require large $l\sub{mr}$. These include (Al,Ga)As heterostructures \cite{deJong, Molenkamp, ScanningGateGaAs}, GaAs \cite{ViscousFlow2DGas}, PdCoO$_2$ crystals \cite{PdCoO2}, WP$_2$ \cite{WP2} and graphene \cite{NegativeLocal, SuperBallisticExpt, FluidityOnsetExpt, GrapheneMHD}.

Signatures and novel consequences \cite{LF2016, LF2017, Torre2015, WhirlpoolsOrNot, Gurzhi, NegativeMagnetoresistance, StokesParadox1, StokesParadox2, ViolationWF, HydroHallViscosity1, HydroHallViscosity2, HydroHallViscosity3, PhononEmission, SuperBallisticTheory, FermiJets, Tomadin2013, CorbinoDisc, SoundModes, FalkovichAC, DyaShu, FluidityOnsetTheory, HydroElectronHolePlasma, HydroToBallisticCrossoverDirac, PreTurbulenceProspects} of a hydrodynamic regime in a Fermi liquid (also referred to as a viscous regime) have recently been the focus of intense activity, since the regime arises from enhanced MC interactions. Calculations based on fluid models indicate that DC charge transport by viscous electrons obeys a nonlocal current-voltage relation and produces a \emph{negative} nonlocal resistance \cite{LF2016, Torre2015}, in sharp contrast to the local current-voltage relation of the Ohmic regime; clearly, interactions do not simply renormalize the conductivity. Perhaps even more striking is the possibility of generating current vortices in a hydrodynamic flow \cite{LF2016, LF2017, Torre2015, WhirlpoolsOrNot}, which are absent in an Ohmic regime. 

Distinct from the hydrodynamic regime is the ballistic regime, which arises when both MR and MC interactions are rare. The absence of MR interactions implies that this regime \emph{also} conserves momentum in the bulk. This results in a degeneracy in DC transport between the hydrodynamic and ballistic regimes. We show that \emph{both} have negative resistances (fig.~2c) and, remarkably, even current flows ordered into vortices (fig.~3). Thus, a negative resistance only indicates a non-Ohmic regime (hydrodynamic/ballistic) resulting from the \emph{absence} of MR interactions, \emph{not} the dominance of bulk MC interactions. Experiments \cite{SuperBallisticExpt,NegativeLocal, PdCoO2,deJong, Molenkamp, FluidityOnsetExpt} currently resolve the hydrodynamic-ballistic degeneracy by changing temperature, carrier concentration or device/contact geometry, and fitting against an expected hydrodynamic or ballistic response to the altered conditions\cite{Gurzhi, SuperBallisticTheory,  StokesParadox1, StokesParadox2,   FluidityOnsetTheory}.

In this paper, we introduce and exploit AC transport as a powerful technique for studying non-Ohmic regimes. We show that each of hydrodynamic, ballistic and Ohmic regimes can be \emph{directly} identified using spatiotemporal correlations in AC transport without \emph{any} change in experimental conditions. Further, we show that vortex formation can be accessed in AC transport with much less fine-tuning compared to DC transport. A snapshot of the key results is presented in fig.~1.

We illustrate each of these points by considering graphene (for concreteness) in a generic device geometry. We show that DC sources fail to generate vortices in the presence of even a modest amount of MR scattering ($\tau\sub{mr}\sim 10$ ps), and show that these are easily generated by switching to AC sources of experimentally accessible frequencies ($\sim$ GHz). In fact, vortex dynamics are crucial for AC transport to proceed both in the hydrodynamic and ballistic regimes. The associated current reversals occur via repeated vortex formation and mergers through reconnection and annihilation; mechanisms seen widely in classical two-dimensional fluids.

We obtain our results by solving for the dynamics of quasiparticles using a deterministic high-resolution numerical scheme which converges much faster than statistical particle methods. The kinetic approach naturally gives rise to all three transport regimes by simply varying $\tau\sub{mr}$ and $\tau\sub{mc}$. In particular, the ballistic regime cannot be accessed by effective fluid models and requires solving for the full time-dependent non-equilibrium distribution function over the entire device. This is made possible using an efficient implementation of our numerical scheme in the package {\tt bolt} \cite{bolt} that exploits the massive computational power of GPU computing clusters.

The paper is structured as follows. In \S\ref{sec:model}, we describe the kinetic model which we solve for the graphene device outlined in \S\ref{sec:setup}. We then study DC transport for this device in \S\ref{sec:DC} and highlight the challenges of using DC signatures to discern the different regimes. Next, we proceed to AC transport and present vortex dynamics and the distinct spatiotemporal correlations of the regimes in \S\ref{sec:AC}. In \S\ref{sec:phase_diagram}, we exploit these correlations to map all the regimes in the $\{\tau\sub{mc}, \tau\sub{mr}\}$ phase space for the device and contact geometry considered. Finally, we provide an argument in \S\ref{sec:vortex_ordering} to understand why vortex ordering occurs in both the hydrodynamic and ballistic regimes and then conclude in \S\ref{sec:conclusion}. Additionally, in appendix \S\ref{sec:appendix_A}, we show how our kinetic calculations incorporate the effect of long-range Coulomb interactions and also show how they appear in fluid models in appendix \S\ref{sec:appendix_B}. The appendices also discuss two different approximations used to compute current-voltage characteristics and verify that they are consistent with each other.

\begin{figure*}[!htbp]\label{fig:summary}
\begin{center}
\includegraphics[width=190mm]{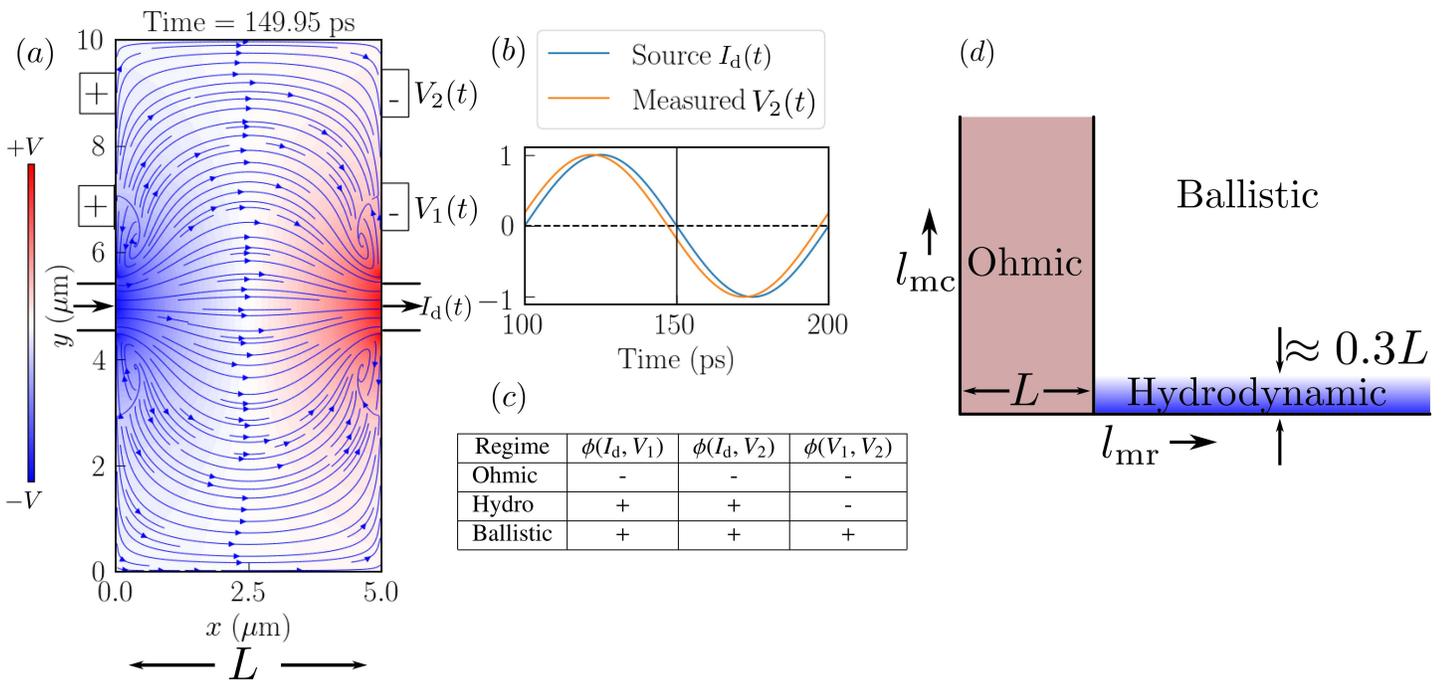}
\end{center}
\caption{\emph{Summary}: (a) Current streamlines and potential in a graphene device with dimensions $5$ $\mu$m $\times$ 10 $\mu$m connected to a 10 GHz AC source. The momentum-conserving and momentum-relaxing time-scales are $\{\tau\sub{mc}, \tau\sub{mr}\} = \{0.2, 5\}$ ps. Vortices are generated near the driving leads ($y = [4.5, 5.5]\;\mu$m on the left and right edges), close to the half cycle of the source ($t=150$ ps). In contrast, there are no vortices and no regions of negative resistance in DC transport for the same set of parameters (fig.~2c). (b) The normalized time series of the current source $I(t)$ and the measured voltage $V_2(t)$. The vertical line indicates the time at which the snapshot in (a) is shown. The measurement leads the source producing a $\phi(I\sub{d}, V_2)>0$. (c) Summary of the correlation signatures of all the regimes. Identifying a regime uniquely requires two voltage measurements as shown by the placement of probes in (a). However, a single measurement anywhere on the edge can distinguish Ohmic and non-Ohmic regimes. (d) Regime boundaries for the device geometry shown in (a), identified using the signatures shown in (c). The hydrodynamic regime requires $l\sub{mr} \gtrsim L$ \emph{and} $l\sub{mc} \lesssim 0.3 L$, where $L$ is the width of the device.}
\end{figure*}

\section{Kinetic Model} \label{sec:model}

We consider graphene (a) well above the charge neutrality point, where quasiparticle excitations are well-defined, and (b) over length scales ($\sim \mu m$) where quantum interference effects are washed out. Transport is then described by the Boltzmann equation that governs the evolution of a charge carrier distribution $f(\mathbf{x}, \mathbf{p}, t)$ in the 4-dimensional phase space of spatial $\mathbf{x} \equiv (x, y)$ and momentum $\mathbf{p} \equiv (p\sub{x}, p\sub{y})$ coordinates,
\begin{align}
\frac{\partial f}{\partial t} +  \mathbf{v}\cdot\frac{\partial f}{\partial \mathbf{x}} & = - \frac{f - f\super{mr}_{0} }{\tau\sub{mr}} -  \frac{f - f\super{mc}_0 }{\tau\sub{mc}}\label{eqn:fe_evol}
\end{align}
where the velocity $\mathbf{v} = \partial \mathcal{E}/\partial \mathbf{p}$ and $\mathcal{E}(\mathbf{p})$ is the band energy. For doped graphene in the upper band, $\mathcal{E}(p) = v\sub{F}\; p \implies \mathbf{v} = v\sub{F}\; \hat{\mathbf{p}}$, where $v\sub{F} \approx 10^6$ m/s = 1 $\mu$m/ps is the Fermi velocity. The terms on the right are the MR and MC collision operators, parametrized in a relaxation time approximation by $\tau\sub{mr}$ and $\tau\sub{mc}$ respectively. In writing (\ref{eqn:fe_evol}), we are working in the $\mu/(k_BT) \gg 1$ regime where the lower band is not needed. 

Electrical transport is set up through current injection at device boundaries which gives rise to a chemical potential gradient. We show in appendix \S\ref{sec:appendix_A} that backreaction $-e\mathbf{E}\cdot \partial f/\partial \mathbf{p}$ ($e\equiv$ charge magnitude) of the self-consistent electric fields $\mathbf{E}$ is incorporated at linear order, even though the term is not explicitly present in (\ref{eqn:fe_evol}).

The collision operators relax $f(\mathbf{x}, \mathbf{p}, t)$ to stationary and drifting local Fermi-Dirac distributions, $f\super{mr}_0(\mu\sub{mr}, T\sub{mr})$ and $f\super{mc}_0(\mu\sub{mc} + \mathbf{p}\cdot \mathbf{v}\sub{d}, T\sub{mc})$ respectively. The spatiotemporal Lagrange multipliers $\{\mu\sub{mr}(\mathbf{x}, t), \mu\sub{mc}(\mathbf{x}, t)\}$, $\{T\sub{mr}(\mathbf{x}, t), T\sub{mc}(\mathbf{x}, t)\}$ and $\mathbf{v}\sub{d}(\mathbf{x}, t)$ are needed for charge, energy and momentum conservation respectively. These are solved for by imposing the matching conditions, 
\begin{align}
\langle f\super{mr}_0 \rangle = \langle f \rangle = \langle f\super{mc}_0 \rangle \label{eqn:mass_cons}\\
\langle \mathcal{E}(p) f\super{mr}_0 \rangle = \langle \mathcal{E}(p) f \rangle = \langle  \mathcal{E}(p) f\super{mc}_0 \rangle \label{eqn:energy_cons}\\
\langle\mathbf{p}f\rangle = \langle\mathbf{p}f\super{mc}_0\rangle \label{eqn:mom_cons}
\end{align}
where $\langle\rangle = N/(2 \pi \hbar)^2 \int d^2p$ and $N = 4$.

The model thus evolves the four-dimensional electron distribution function $f(\mathbf{x}, \mathbf{p}, t)$ by additionally solving the six Lagrange multiplier constraints (\ref{eqn:mass_cons}, \ref{eqn:energy_cons}, \ref{eqn:mom_cons}) at every time step. The free parameters in the model are ${\tau\sub{mc}, \tau\sub{mr}}$. We treat these as numerical inputs in units of picoseconds and do not invoke any functional dependence on thermodynamics quantities such as temperature and number density. However, the model incorporates temperature smearing of the Fermi surface, although we find that all results are independent of this effect. We numerically integrate this computationally expensive system on a GPU cluster using {\tt bolt} \cite{bolt}, a fast, massively parallel high-resolution solver for kinetic theories based on a finite volume method to achieve $\mathcal{O}(\Delta x^2, \Delta p^2, \Delta t^2)$ accuracy, where $\Delta x$, $\Delta p$ and $\Delta t$ are the sizes of discrete elements in real space, momentum space and time respectively.

\subsection{Current-Voltage Relationship} \label{sec:IV}

A key quantity of interest is the  current-voltage relationship. After $f(\mathbf{x}, \mathbf{p}, t)$ has been solved for, the current $\mathbf{j}(\mathbf{x}, t)$ is easily computed using $\mathbf{j}(\mathbf{x}, t) = -e\langle \mathbf{p} f \rangle$. However, to obtain the voltage $V(\mathbf{x}, z, t)$, one needs to solve the 3D Poisson equation,
\begin{align} \label{eqn:poisson}
\nabla\cdot\left(\epsilon\sub{r} \nabla V\right) = 4\pi e n \delta(z - z_0)
\end{align}
where $n(\mathbf{x}, t) = \langle f \rangle$ is the 2D charge carrier density, $z_0$ is the location of the 2D sample in the perpendicular direction and $\epsilon\sub{r}(\mathbf{x})$ is the dielectric function. 

To proceed, we assume a graphene field-effect transistor geometry with a dielectric substrate $\epsilon\sub{r}$ of thickness $d \ll L$, where $L \sim \mu$m is a lateral device scale. The mean carrier density $n_0$ can be set to a desired level by applying a backgate voltage $V(\mathbf{x}, z=0) = V\sub{g}$ on the substrate (located at $z=0$) beneath the dielectric. This is given by,
\begin{align} \label{eqn:LCA_original}
n_0 = \frac{C V\sub{g}}{e}
\end{align}
where $C = \epsilon\sub{r}/(4\pi d)$ is the capacitance per unit area.

Further, we require the relationship between charge inhomogeneities and \emph{in-plane} voltage fluctuations. For the field-effect geometry, an analytic solution of the 3D Poisson equation in the 2D plane of the device has been computed by Tomadin and Polini \cite{Tomadin2013}. To $\mathcal{O}(d/L)$, this solution simplifies to
\begin{align} \label{eqn:LCA}
\Delta V \approx -\frac{e}{C} \Delta n
\end{align}
where $\Delta V$ and $\Delta n$ are the in-plane voltage and charge density differences respectively.  Note that while (\ref{eqn:LCA_original}) and (\ref{eqn:LCA}) appear similar, they relate different quantities. The approximation (\ref{eqn:LCA}), referred to as ``gradual channel'' in \cite{Tomadin2013} and ``local capacitance'' in \cite{CorbinoDisc}, thus relates $f$ obtained by solving (\ref{eqn:fe_evol}) to the voltages measured. As explained in appendix \S\ref{sec:appendix_A}, this computation is a post-processing step and is not required for the evolution of (\ref{eqn:fe_evol}) to linear order.

\begin{figure}[!htbp]\label{fig:DC}
\begin{center}
\includegraphics[width=\columnwidth]{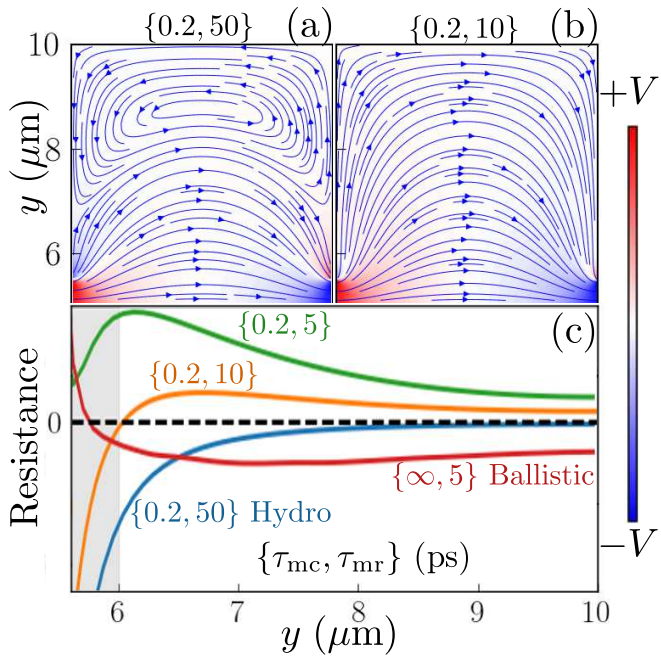}
\end{center}
\caption{\emph{DC transport} (\S\ref{sec:DC}): Current streamlines and potentials in the symmetric top half of the device shown in fig.~1a for $\{\tau\sub{mc},\tau\sub{mr}\} = $ (a) $\{0.2, 50\}$ ps, (b) $\{0.2, 10\}$ ps. (c) Resistance $V(y)/I$, where $V(y)$ is the potential for y $\in (5.5, 10]\;\mu$m and $I$ is the injected current through contacts between $y = [4.5, 5.5]$ $\mu$m. The negative resistance for $\tau\sub{mc} = 0.2$ ps (fast MC interactions) is nonlocal for (a) $\tau\sub{mr}=50$ ps, becomes local for (b) $\tau\sub{mr}=10$ ps (shaded region $\approx 0.5\;\mu$m)and disappears everywhere for $\tau\sub{mr}=5$ ps. All these cases are hydrodynamic in AC transport. The ballistic regime (zero MC interactions, red) is degenerate with hydro (blue); it also has a negative resistance and a flow profile similar to (a) (shown in fig.~3)}
\end{figure}

\begin{figure}[!htbp]\label{fig:S1_ohmic_vs_ballistic}
\begin{center}
\includegraphics[width=\columnwidth]{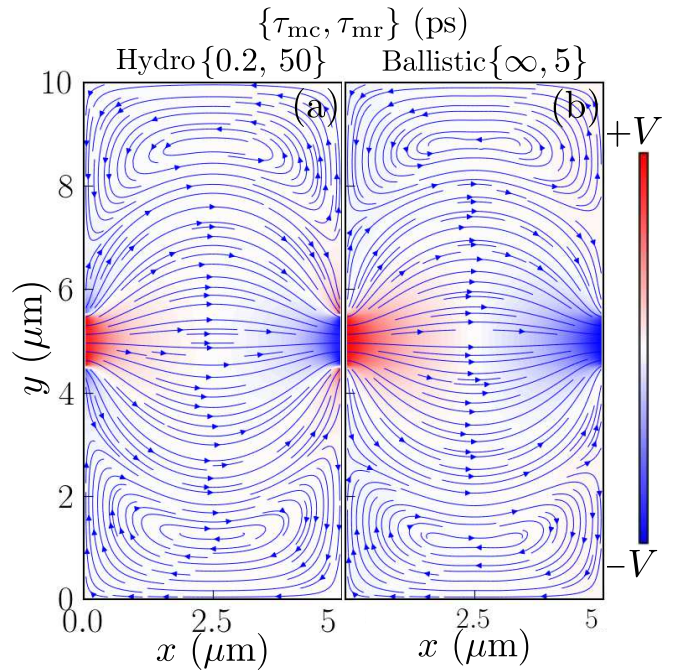}
\end{center}
\caption{\emph{Hydro-Ballistic degeneracy} (\S\ref{sec:DC_hydro_ballistic}): Current streamlines and potentials in DC transport for (a) hydro regime with $\{\tau\sub{mc}, \tau\sub{mr}\} = \{0.2, 50\}$ ps and (b) ballistic regime with $\{\tau\sub{mc}, \tau\sub{mr}\} = \{\infty, 5\}$. The flow profiles are strikingly similar and the nonlocal resistance computed using voltage measured along the edge is negative for both cases (see fig.~2c).}
\end{figure}

\begin{figure*}[!htbp]\label{fig:movie}
\begin{center}
\includegraphics[width=180mm]{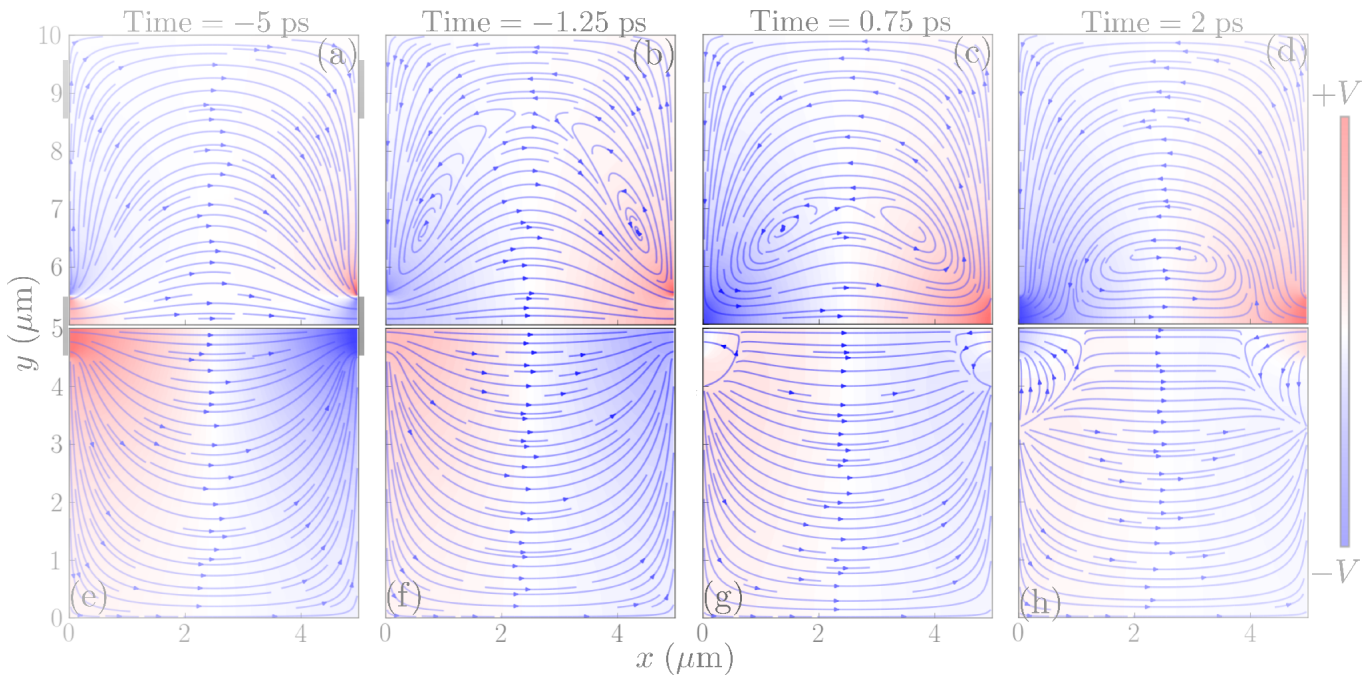}
\end{center}
\caption{\emph{Vortex dynamics} (\S\ref{sec:vortex_dynamics}): Current streamlines and potentials. (a-d) Time evolution of a hydrodynamic mode excited by a 10 GHz AC source for $\{\tau\sub{mc}, \tau\sub{mr}\}=\{0.2, 10\}$ ps (same parameters as fig.~2b), through contacts between $y = [4.5, 5.5]\;\mu$m. The device is reflection symmetric about its center and we present the top half. The source reverses at $t = 0$ ps. At (a) $t=-5$ ps, the voltage everywhere along the edge goes against the source. Evolution proceeds through (b) vortex generation, (c,d) merger through reconnection. Note that the voltage in the entire device has changed sign at $t<0$ ps. In contrast, AC transport in an Ohmic regime (e-h), shown here for $\{\tau\sub{mc}, \tau\sub{mr}\}=\{0.2, 1\}$ ps, proceeds by (g) wave-fronts that originate at the source and (h) propagate into the device. See also movies for hydrodynamic\footnote{\url{https://vimeo.com/261891439}}\& Ohmic \footnote{\url{https://vimeo.com/261891102}} regimes.}
\end{figure*}

\section{Setup} \label{sec:setup}

We consider a 5 $\mu$m $\times$ 10 $\mu$m graphene device, with drive contacts $1\;\mu m$ wide at the center of the left and right edges (see fig.~1a). We assume an electron density of $n = 10^{12}$ cm$^{-2}$ , and ideal Ohmic contacts i.e., the Fermi level of the contact metal is the same as the electron chemical potential in graphene (at the chosen carrier density). 
At the contacts, we impose Dirichlet boundary conditions that implement a current source/sink, with the distribution at both contacts set to a shifted Fermi-Dirac ($f\super{mc}_0$) with the drift velocities $\mathbf{v}\super{L}\sub{d} = \mathbf{v}\super{R}\sub{d} = (v(t), 0)$, where L and R denote the left and right contacts respectively and $v(t)$ is a time-dependent magnitude. For DC calculations, we set $v(t) = v_0 = 10^{-4}\; v\sub{F}$, corresponding to a current injection of $\sim 0.1\;\mu A$. For AC calculations, we set $v(t) = v_0 \sin(2 \pi f t)$, where $f$ is the source frequency. We present results for $f = 10$ GHz, but they are valid over a wide range of frequencies as discussed later.

On the device boundary outside of the contacts, we impose perfect reflection on the electrons (specular scattering). This corresponds to ``free-slip'' boundaries in the parlance of fluid models, as opposed to ``no-slip" boundaries. The question of what boundary conditions are correct is an open one, but there is increasing evidence in support of free-slip boundaries because of the suppression of the Gurzhi effect \cite{Gurzhi} in graphene, which becomes dominant only in the presence of no-slip boundaries \cite{Torre2015, NegativeLocal}.

\section{DC transport} \label{sec:DC}

We first examine signatures of a hydro regime in DC transport. For fast MC and very slow MR interactions, $\{\tau\sub{mc}, \tau\sub{mr}\}$ = $\{0.2, 50\}$ ps $\implies \tau\sub{mr}/\tau\sub{mc} = 250$, we get current vortices flowing against the drive (fig.~2a); the distinctive features of a fluid. Further, the nonlocal resistance computed using voltage measured far from the driving leads divided by the injected current is \emph{negative} (fig.~2c), as shown using fluid models \cite{LF2016, Torre2015, WhirlpoolsOrNot, LF2017}. Now consider $\{\tau\sub{mc}, \tau\sub{mr}\}$ = $\{0.2, 10\}$ ps $\implies \tau\sub{mr}/\tau\sub{mc} = 50$, still expected to be deep in the hydro regime. However, the vortices no longer appear (fig.~2b). We have verified their absence everywhere in the domain down to 25 nm ($\sim 0.1 \times$ e-e
mean free path); well below the length-scale at which a
hydrodynamic description is expected to apply. In addition, the nonlocal resistance is negative
only locally, within $0.5\; \mu$m of the driving leads (fig.~2(b,c)). A further reduction in $\tau\sub{mr}$ to 5 ps ($\tau\sub{mr}/\tau\sub{mc}=25$) leads to a disappearance of this negative resistance as well (fig.~2c). 

\subsection{Hydro-Ballistic Degeneracy} \label{sec:DC_hydro_ballistic}

Consider now the ballistic regime set by the parameters $\{\tau\sub{mc} = \infty, \tau\sub{mr} \simeq L/v_F = 5\}$ ps, i.e. zero MC interactions and weak MR interactions. The ballistic flow is qualitatively indistinguishable from a hydrodynamic flow (both shown in fig.~3), \emph{and} produces a negative resistance all along the edge (fig.~2c). 

Thus, we see that not only are DC sources inefficient at exciting hydrodynamic behaviour in parameter regimes that are clearly MC dominated but also fail to distinguish between ballistic and hydrodynamic regimes. Note that the specific thresholds between regimes we report here are
for the device and contact geometry shown in fig.~1a, and these would vary for other geometries. In particular, the geometry proposed by Torre {\it et al.}~\cite{Torre2015} to investigate hydrodynamic effects has contacts on the same side of the device and is much more conducive to vortex generation in DC transport \cite{WhirlpoolsOrNot}. Our choice of geometry, adopted from Levitov \& Falkovich \cite{LF2016}, has been made so as to demonstrate vigorous vortex production even in the least favorable setup, as we shall see in the next section.

\begin{figure}[!htbp]\label{fig:time_series}
\begin{center}
\includegraphics[width=\columnwidth]{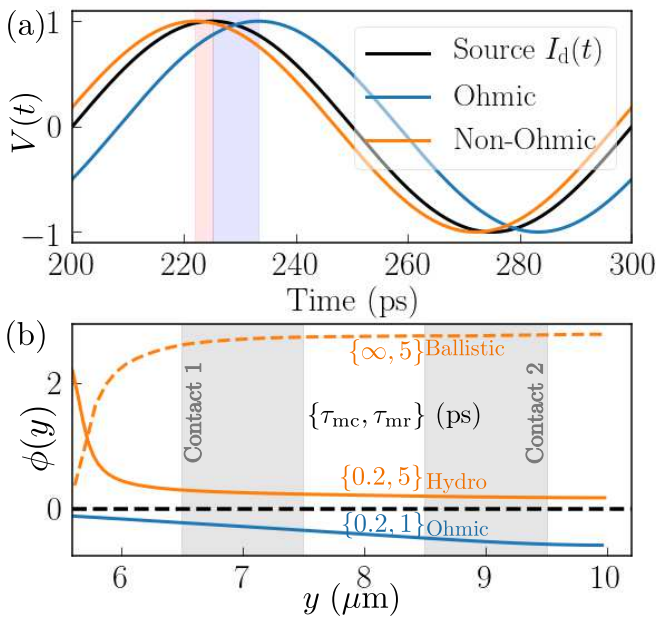}
\end{center}
\caption{\emph{Spatiotemporal correlations} (\S\ref{sec:correlations}): (a) The normalized voltages $V(t)$ measured by a 1 $\mu$m contact between $y = [8.5, 9.5]\;\mu$m. Note that all the curves are sinusoidal with the source frequency of 10 GHz and differ only by a well-defined phase $\phi(I\sub{d}, V)$. An Ohmic regime (here $\{\tau\sub{mc}, \tau\sub{mr}\} = \{0.2, 1\}$ ps) produces a time-series that \emph{lags} the source (shaded in blue). However, the signal in non-Ohmic regimes (shown for $\{0.2, 5\}$ ps) \emph{lead} the source (shaded in red). (b) $\phi(I\sub{d}, V)$ vs distance from the driving leads. The sign of $\phi$ is spatially highly extended (contrast with fig.~2(c)). While both hydro and ballistic regimes have $\phi > 0$, the slope in hydro is negative whereas it is positive in ballistic. This is measured using the two-point correlation $\phi(V_1, V_2) = \phi(I\sub{d}, V_2) - \phi(I\sub{d}, V_1)$. The shaded region shows the locations and widths of contacts used to make the phase diagrams fig.~(6,7).
}
\end{figure}

\section{AC transport} \label{sec:AC}

The situation improves dramatically if we use an AC source with frequency $f \ll v\sub{F}/L = 200$ GHz. As long as $l\sub{mr} \equiv \tau\sub{mr}v\sub{F} \gtrsim L$, MC scattering (either specular boundary or bulk e-e) excites collective modes involving vortices that are continuously generated and destroyed at the rate $f$.
This condition in AC is far more enabling for vortex generation in the hydrodynamic regime ($l\sub{mc} \ll L$) than in DC which requires $D \sim v\sub{F} \sqrt{\tau\sub{mc}\tau\sub{mr}}/2 \gtrsim L/(\sqrt{2} \pi)$, where $D$ is the vorticity diffusion length \cite{WhirlpoolsOrNot}. Therefore, vortex generation occurs in AC even for parameter regimes where current streamlines appear distinctly Ohmic in DC.
For example, with $\{\tau\sub{mc}, \tau\sub{mr}\}$ = $\{0.2, 10\}$ ps,
$l\sub{mr} \approx 2\cdot L$ enabling vortices in AC as shown in fig.~4(b-d),
but $D \approx 0.6\cdot L/(\sqrt{2}\pi)$ resulting in Ohmic-like DC transport in fig.~2(b).
Indeed, vortices form in AC even for the marginal case of
$\{\tau\sub{mc}, \tau\sub{mr}\} = \{0.2, 5\}$ ps,  where $l\sub{mr} \approx L$
(fig.~1a).

\subsection{Vortex dynamics} \label{sec:vortex_dynamics}

Fig.~4(a-d) show the flow structure of the hydro mode excited by a 10 GHz source for $\{\tau\sub{mc}, \tau\sub{mr}\}$ = $\{0.2, 10\}$ ps. As the source is about to change sign near the half-cycle, vortices form symmetrically at the left and right contacts resulting in a quadrupolar mode (fig.~4b). These vortices, with the same sign of vorticity, grow and merge through reconnection (fig.~4c); as in 2D classical fluids. There is now a dipolar mode in the device with vortices in the top and bottom halves having opposite signs of vorticity (fig.~4d). These then annihilate in the middle, and allow the flow to reverse. In contrast, Ohmic AC transport proceeds through wave-fronts (fig.~4g) that originate from the contacts and travel into the device (fig.~4(h)). We note that ballistic AC transport also proceeds through vortex dynamics, albeit with an altered choreography; vortices form at the top/bottom boundaries and move inwards.

\subsection{Spatiotemporal correlations} \label{sec:correlations}

The persistent time-dependence of an AC source produces several useful spatiotemporal correlations. We consider the phase $\phi(I\sub{d}, V)$ between a nonlocal voltage $V(t)$ measured by contacts on the edge (top of fig.~1a), and the current source $I\sub{d}(t)$ (center of fig.~1a). An Ohmic regime is defined by a local current-voltage relationship; an injected current causes local changes in voltage ($\propto I\sub{d}$) which then propagate into the device (fig.~4(g,h)). Therefore, the measured $V$ \emph{lags} the source $I\sub{d}$ (fig.~5a) and the phase $\phi$ is negative.

A transition from a negative to a positive phase (fig.~5a) signifies the breakdown of a local current-voltage relation and the onset of a \emph{non}-Ohmic regime (fig.~5b) with a \emph{non}-local current-voltage relation (see \cite{LF2016, Torre2015, WhirlpoolsOrNot, LF2017} for hydro).
This arises whenever MR interactions are weak, which is a necessary
but not sufficient condition for a hydrodynamic regime.
Hydrodynamic transport also requires strong MC interactions that impose a local equilibrium
\emph{in the bulk}, unlike specular scattering at the boundary in the ballistic regime.

This key difference in the ballistic regime
is directly captured by the two-point correlation
$\phi(V_1, V_2) = \phi(I\sub{d}, V_2) - \phi(I\sub{d}, V_1)$,
with a $V_1(t)$ measured closer to the source (schematic shown in fig.~1a).
When voltage gradients flow through bulk interactions
(MC or MR), $V_2$ must lag $V_1$ and $\phi(V_1, V_2) < 0$.
This is evident from the negative slope of $\phi(I\sub{d}, V)$
in both Ohmic and hydrodynamic regimes (fig.~5b).
The ballistic regime which only has boundary scattering
exhibits a positive slope in $\phi(I\sub{d}, V)$ (fig.~5b),
so that $\phi(V_1, V_2) > 0$ and $V_2$ closer to the
top boundary now \emph{leads} $V_1$.

\subsection{AC vs DC signatures}

The $\phi(I\sub{d}, V)>0$ diagnostic of non-Ohmicity is similar to negative resistance in DC, with both indicating a nonlocal current-voltage relation. However, $\phi(V_1, V_2)$ has no analogue in DC; the slope of resistance does not show any robust pattern (fig.~2c). In addition, the spatial extent of the phase correlations is much greater in a hydrodynamic regime. For $\{\tau\sub{mc}, \tau\sub{mr}\}$ = $\{0.2, 10\}$ ps, the negative resistance is confined to within $0.5\;\mu$m of the driving leads (fig.~2c). In contrast, the phase at 10 GHz for even $\{0.2, 5\}$ ps is positive over the \emph{entire} edge (fig.~5b), and there is no negative resistance anywhere for these parameters in DC (fig.~2c).

\begin{figure}[!htbp]\label{fig:phase_diff}
\begin{center}
\includegraphics[width=\columnwidth]{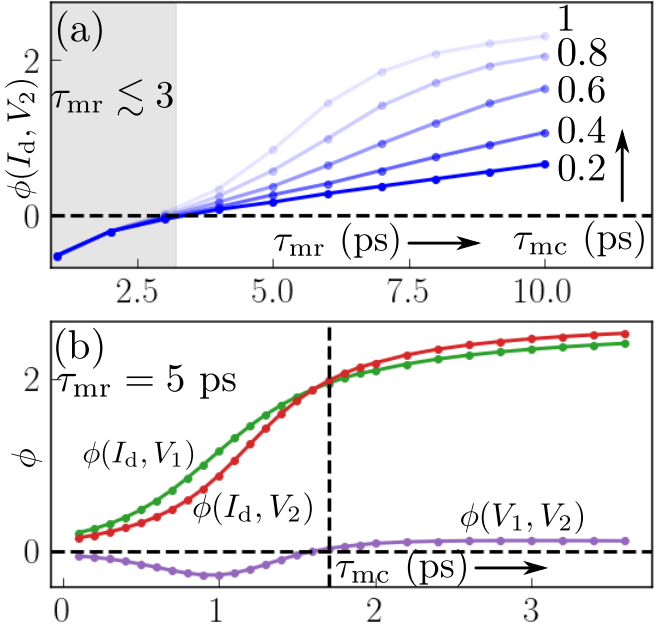}
\end{center}
\caption{\emph{Phase diagram} (\S\ref{sec:phase_diagram}): (a) The phase $\phi(I\sub{d}, V_2)$ measured at 10 GHz using a 1 $\mu$m contact (contact-2 in fig.~3b) on the device shown in fig.~1(a). A non-Ohmic regime corresponds to $\phi > 0$. The condition for non-Ohmic AC transport, $\tau\sub{mr} \gtrsim L/v\sub{F} = 5$ ps, is well-satisfied. The two-point correlation $\phi(V_1, V_2)$ is negative for all parameters here and so this plot shows an Ohmic to hydro transition. The voltage $V_1$ is measured using contact-1 in fig.~3b. (b) Hydro to ballistic transition: On varying $\tau\sub{mc}$ for a fixed $\tau\sub{mr} = 5$ ps, $\phi(V_1, V_2)$ changes sign from negative to positive for $\tau\sub{mc} \gtrsim 1.7$ ps signaling a transition into the ballistic regime. Note that $\phi(I\sub{d}, V_1), \phi(I\sub{d}, V_1) > 0$ throughout.
}
\end{figure}

\section{Phase diagram} \label{sec:phase_diagram}

We now use the phases $\phi_1 \equiv \phi(I\sub{d}, V_1)$, $\phi_2 \equiv \phi(I\sub{d}, V_2)$ and $\phi_{12} \equiv \phi(V_1, V_2)$, measured using contacts shown in fig.~1a, to map all regimes in the $\{\tau\sub{mc}, \tau\sub{mr}\}$ parameter space for our device. We consider a wide range of these parameters, all of which satisfy $\tau\sub{mc}/\tau\sub{mr} < 1$; parameters where the hydrodynamic regime could potentially arise.

Fig.~6a shows $\phi_2>0$ for all $\tau\sub{mr} \gtrsim 3$ ps (same for $\phi_1$, not shown), in accordance with the $\tau\sub{mr}\gtrsim L/v\sub{F} = 5$ ps condition for non-Ohmic AC transport. Further, we find $\phi_{12} < 0$ for all parameters in fig.~6a, indicating an Ohmic to hydrodynamic transition. To now transition from hydrodynamic to ballistic, we vary $\tau\sub{mc}$ for a fixed $\tau\sub{mr}=5$ ps in fig.~6b. We indeed find non-Ohmicity ($\phi_1, \phi_2 > 0$) over the entire range of $\tau\sub{mc}$ \emph{but} bulk scattering ($\phi_{12}<0$) only for $\tau\sub{mc} \lesssim 1.7$ ps $\approx 0.3 L/v\sub{F} $; consistent with the $l\sub{mc}\equiv\tau\sub{mc}v\sub{F} \ll L$ requirement of a hydrodynamic regime. Thus we need \emph{both} $\tau\sub{mr} \gtrsim 3$ ps and $\tau\sub{mc} \lesssim 1.7$ ps, quantifying the weakness and strength of MR and MC interactions needed for a hydrodynamic regime in our device geometry.

Both conditions on MR and MC interactions are within reach of current devices\cite{NegativeLocal, SuperBallisticExpt, FluidityOnsetExpt}, with the requirement of weak MR further mitigated by adjusting the lateral scale such that $L \lesssim l\sub{mr}\equiv\tau\sub{mr}v\sub{F}$. As discussed in \S\ref{sec:AC}, this condition is much weaker than that required for vortices in DC where the the minimum $\tau\sub{mr}$ scales as $L^2$. The critical $\tau\sub{mr}$ in both AC and DC is shown in fig.~7 for a fixed $\tau\sub{mc} = 0.2$ ps. This $\tau\sub{mc}$ is well within the limit $\lesssim 0.3 L/v\sub{F}$ required for a hydrodynamic regime for the smallest device considered ($L = 1.8$ $\mu$m), once the criterion on $\tau\sub{mr}$ is satisfied. The regime boundaries for the device geometry are summarized in fig.~1d. Notably, the parameter space of the ballistic regime is much larger than that of the hydrodynamic regime.

\begin{figure}[!htbp]\label{fig:phi_cutoff}
\begin{center}
\includegraphics[width=\columnwidth]{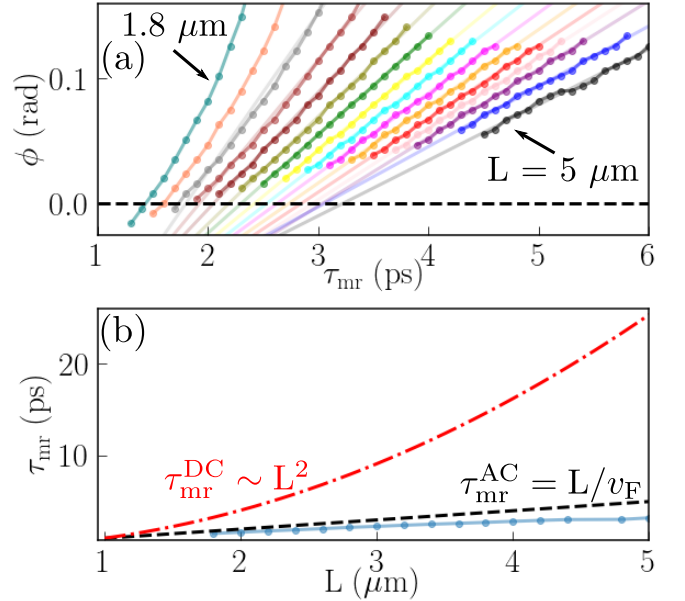}
\end{center}
\caption{\emph{Phase diagram} (\S\ref{sec:phase_diagram}): (a) Phase vs $\tau\sub{mr}$ for devices with varying widths $L \in [1.8, 5]$ $\mu$m in increments of $0.2$ $\mu$m. All the calculations are for a fixed $\tau\sub{mc} = 0.2$ ps, i.e., fast MC interactions. The solid lines are fits to the numerical data and are used to evaluate the critical $\tau\sub{mr}$ for which $\phi(I\sub{d}, V_2)$ changes sign, denoting a transition from an Ohmic to a non-Ohmic regime. The non-Ohmic regime for all cases shown here is hydrodynamic since $\tau\sub{mc} < 0.3 L/v\sub{F}$ for the smallest device considered. (b) The critical $\tau\sub{mr}$ for AC and DC. The critical timescale in AC (blue line) is within the bound $L/v\sub{F}$ whereas it is $[(2/\pi^2)L^2/l\sub{mc}]/v\sub{F}$ in DC \cite{WhirlpoolsOrNot}.}
\end{figure}

\begin{figure}[!htbp]\label{fig:ohmic_to_ballistic}
\begin{center}
\includegraphics[width=\columnwidth]{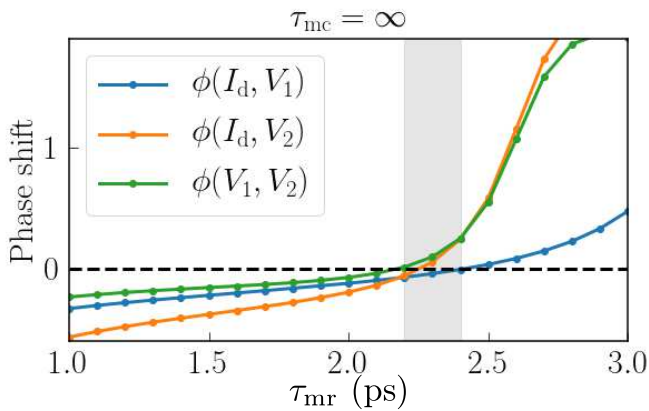}
\end{center}
\caption{\emph{Consistency check} (\S\ref{sec:consistency_check}): For the following calculation, we set $\tau\sub{mc} = \infty$. In an Ohmic regime, all of $\phi(I\sub{d}, V_1)$, $\phi(I\sub{d}, V_2)$ and $\phi(V_1, V_2) = \phi(I\sub{d}, V_2) - \phi(I\sub{d}, V_1)$ are negative, whereas they are all positive in the ballistic regime. Therefore, consistency within the correlations requires that as $\tau\sub{mr}$ is increased to transition from an Ohmic to the ballistic regime, \emph{each} of $\phi(I\sub{d}, V_1)$, $\phi(I\sub{d}, V_2)$  and $\phi(V_1, V_2)$ should change sign from negative to positive at the \emph{same} $\tau\sub{mr}$. The above plot shows that this is indeed the case with a transition at $\tau\sub{mr} = 2.3 \pm 0.1$ ps.}
\end{figure}

\subsection{Consistency check} \label{sec:consistency_check}

We additionally consider the transition from Ohmic to ballistic by setting $\tau\sub{mc} = \infty$ and varying $\tau\sub{mr}$. A correct identification of this transition requires a deep consistency within the correlations; \emph{each} of $\phi_1$, $\phi_2$ and $\phi_{12}$ must change sign from negative to positive at the \emph{same} $\tau\sub{mr}$. Fig.~8 of the phases versus $\tau\sub{mr}$ indeed shows this at $\tau\sub{mr} = 2.3 \pm 0.1$ ps, with the small spread due to finite width (1 $\mu$m) contacts placed a finite distance (1 $\mu$m) apart in our setup.

\section{Vortices in Hydro and Ballistic Regimes} \label{sec:vortex_ordering}
The hydro regime is interaction dominated whereas the ballistic regime occurs in the near \emph{absence} of interactions. However, the current flow patterns in both these regimes are strikingly similar. Both regimes have flows organized into vortices, and therefore have a nonzero vorticity ($\omega \equiv \nabla \times \mathbf{v}\sub{d}$). Note that the vorticity is identically zero in the Ohmic regime since $\mathbf{v}\sub{d} \propto \nabla V$.

Let us ask the question, can we deduce flow ordering in both hydrodynamic and ballistic regimes \emph{without} solving the governing equation (\ref{eqn:fe_evol})? We provide here an argument using Landau theory that the ordering principle for the flow is the same in both regimes.

To proceed, we require an energy functional constrained by the symmetries of the system. Consider no MR interactions ($\tau\sub{mr} = \infty$) and fast MC interactions resulting in vanishing electron viscosity $\nu \sim \tau\sub{mc} \rightarrow 0$. We can then invoke a well-known result that applies to inviscid 2D fluids, the conservation of ``enstrophy'',
\begin{align} \label{eqn:enstrophy}
F[\mathbf{v}\sub{d}] = \int d^2r |\nabla \times \mathbf{v}\sub{d}|^2
\end{align}
where $\mathbf{v}\sub{d}$ is the local carrier drift velocity. The conservation of enstrophy follows from momentum conservation of a fluid element \emph{in} two-dimensions \cite{Enstrophy}.

We interpret $F[\mathbf{v}\sub{d}]$ as a Landau energy functional with a local order parameter $\omega$. A vortex ordered state then has $F > 0$ with $F = 0$ denoting no ordering. A variation of $F$ with respect to the velocity field $\mathbf{v}\sub{d}$ gives,
\begin{align}
\delta F = \oint d\mathbf{s}\cdot (\delta \mathbf{v}\sub{d} \times \omega) + \int d^2r\;\delta \mathbf{v}\sub{d}\cdot(\nabla\times \omega)
\end{align}
The first term on the right vanishes with $\delta \mathbf{v}\sub{d} = 0$ on the boundaries, corresponding to the imposition of boundary conditions. Demanding that the variation of $F$ vanish,
\begin{align} \label{eqn:variation}
\delta F = 0 & \implies \nabla \times \omega = 0
\end{align}
Now applying the curl operator on (\ref{eqn:variation}),
\begin{align} \label{eqn:hydro_vorticity}
\nabla^2 \omega = 0
\end{align}
This is in fact the time-independent fluid momentum conservation equation (\ref{eqn:momentum_cty_steady_state}) in the limit $\tau\sub{mr} \rightarrow \infty$, after an application of the curl operator. Clearly, (\ref{eqn:hydro_vorticity}) admits solutions with $F > 0$ depending on the boundary conditions, i.e, device and contact geometry. Therefore, we see that the conservation of enstrophy (\ref{eqn:enstrophy}) allows for flow ordering in the form of vortices.

Now we note that the enstrophy (\ref{eqn:enstrophy}) is \emph{also} conserved in the ballistic regime and thus we expect the same ordering in this regime as well. Indeed, the ballistic regime preserves an infinite number of invariants because the momentum of every carrier is conserved, thus conserving the momentum of a macroscopic fluid element. In contrast, the hydro regime \emph{only} conserves the momentum of the fluid element; the momenta of individual carriers is thermalized by MC interactions.

The conservation of enstrophy (\ref{eqn:enstrophy}) implies that any local vorticity injected from the boundaries \emph{persists} in the device. The device and contact geometry adopted here is an example where vorticity is zero globally but nonzero locally, near the drive contacts. Since the invariant measure (\ref{eqn:enstrophy}) is positive definite, this geometry allows for vortex formation. A counterexample is a wire geometry where the injected vorticity is exactly zero. Note that the arguments presented in this section only show that flow ordering is possible provided a conducive geometry and do not inform on the interaction timescales required.

\section{Discussion and Conclusion} \label{sec:conclusion}

The hydrodynamic transport regime in Fermi liquids has long been theorized, but its robustness has always been in question. In particular, the specific requirements on defect densities, and e-ph, e-e coupling needed to access this regime in various device geometries have been open issues. Indeed, we find that the regime {\em is} very delicate in DC transport, requiring unreasonably low momentum-relaxing scattering or specially designed geometries in order for its effects to manifest themselves. An even bigger concern is its high degeneracy with the ballistic regime in which we find nearly identical current vortex structures and negative resistances.

We have shown that both issues can be resolved by switching to AC transport. \underline{First}, the requirement on momentum-relaxing scattering for the hydrodynamic regime is greatly mitigated, with the criterion simply being $l\sub{mr} \gtrsim L$ as compared to $l\sub{mr} \gtrsim (0.2/l\sub{mc}) L^2$ in DC. Hydrodynamic modes rich in vortices are therefore easily excited in AC transport, with current reversals occurring through intricate vortex dynamics. Note that the condition on $l\sub{mr}$ in AC is independent of $l\sub{mc}$ whereas it becomes increasingly stringent in DC as $l\sub{mc}\rightarrow 0$!

\underline{Second}, there exist strong spatiotemporal correlations in AC transport that are able to uniquely and directly identify all transport regimes (as summarized in fig.~1c). One of these correlations, the phase between a nonlocal voltage and the current drive tests for the locality of the current-voltage relation and can thus differentiate between Ohmic and non-Ohmic regimes. Further, the phase between two nonlocal voltages tests for the presence of bulk interactions and can therefore distinguish between the hydrodynamic and ballistic regimes. These correlations extend over the entire device, allowing for great flexibility in selecting the positions of probe contacts in experiments. In contrast, we have shown that negative resistances in hydrodynamic DC transport are tightly localized to the immediate vicinity of the drive contacts (in the narrow parameter range where they appear at all).

Using the correlation signatures, we have mapped out the regime boundaries in AC transport (shown in fig.~1d) for the graphene device depicted in fig.~1a. The hydrodynamic regime emerges for $l\sub{mr} \gtrsim L$ and $l\sub{mc} \lesssim 0.3L$, with the region $l\sub{mc} > 0.3 L$ occupied by the ballistic regime. Transport for all $l\sub{mr} \gtrsim L$, i.e., both the hydrodynamic and ballistic regimes, occurs through vortex dynamics, a reflection of the high degree of degeneracy between the two regimes. However, the dynamics have different choreographies: For the device shown, the vortices in the hydrodynamic regime form at the drive contacts, whereas they form at the top and bottom boundaries in the ballistic regime. We emphasize here that the voltage-voltage correlation being used to discern the two regimes probes the nature of signal propagation within the device and is largely independent of the dynamics of vortices.

A surprising outcome of our calculations is the presence of current vortices in the ballistic regime; structures which are typically associated with a fluid. We trace this back to the fact that both non-Ohmic regimes conserve momentum and present an argument using Landau theory that vortices arise directly as a consequence of \emph{enstrophy} conservation, which in turn results from momentum conservation in two-dimensions.

Our results have been obtained using {\tt bolt} \cite{bolt}, a package we have developed to directly solve for quasiparticle transport at the kinetic level with band structures and collision operators as inputs. The kinetic framework is crucial to access the ballistic regime, which is not amenable to effective fluid models. The package uses a high-resolution numerical scheme that deterministically discretizes the four-dimensional phase space of 2D systems. While the computational complexity involved {\em is} massive, it is nevertheless within reach of present day GPUs. The package exploits these with sufficient efficiency so as to allow time-dependent device simulations to be routinely carried out in a matter of hours, with realistic device dimensions and interaction time scales.

Finally, while our calculations were performed for a Fermi liquid, our phase diagnostics in AC transport may also hold for strongly interacting quantum systems that admit a hydrodynamic description \cite{HydroStronglyCorrelated}, such as the Dirac fluid which arises at charge neutrality in graphene \cite{GraphenePerfectFluid, DiracFluidTheory, DiracFluid, GrapheneQuantumCriticalConductivity}. Specifically, the current-voltage phase correlation is sensitive to \emph{any} nonlocality in the current-voltage relation, regardless of its precise form. So it will immediately detect the presence of a hydrodynamic regime, whether this originates from a collection of rapidly interacting quasiparticles or in a strongly interacting system devoid of quasiparticles. Similarly, the voltage-voltage phase correlation can be used to ensure the dominance of interactions and rule out ballistic effects.

\appendix
\section{Long-range Coulomb interactions} \label{sec:appendix_A}
In writing the governing equation (\ref{eqn:fe_evol}), we have omitted the term $-e\mathbf{E}\cdot\partial f/\partial\mathbf{p}$ on the left hand side. This term incorporates backreaction of the self-consistent long-range Coulomb interactions, and couples the Boltzmann equation (\ref{eqn:fe_evol}) to the 3D Poisson equation (\ref{eqn:poisson}). Here we show that the dynamical effects of this term are accounted for at linear order through a renormalized chemical potential.

We begin by assuming that the system is driven at constant temperature using a small background chemical potential gradient $\nabla \mu \equiv (\partial \mu/\partial x, \partial \mu/\partial y) \sim \epsilon (\mu\sub{F}/L)$, where $\mu\sub{F}$ is the Fermi level, $L$ is the device scale and $\epsilon \ll 1$ is a dimensionless bookkeeping parameter. This gradient results in a distribution $f = f_0(\mathcal{E}, \mu, T) + \delta f$, where $f_0$ is the background \emph{local} equilibrium distribution and $\delta f \sim \mathcal{O}(\epsilon)$ is a small perturbation. We then have the force due to a self-consistent electric field in the plane of the graphene strip, $-e\mathbf{E}(\mathbf{x}) = e\nabla V(\mathbf{x}, z=z_0) \sim \mathcal{O}(\epsilon)$, where $z=z_0$ is the vertical location of the strip. Substituting $f$ in (\ref{eqn:fe_evol}), now with $-e\mathbf{E}\cdot\partial f/\partial\mathbf{p}$ included, the LHS is
\begin{align} \label{eqn:delta_f}
\frac{\partial \delta f}{\partial t} + \mathbf{v}\cdot\frac{\partial \delta f}{\partial \mathbf{x}} - \left(e\mathbf{E} + \nabla\mu\right)\cdot\frac{\partial f_0}{\partial \mathbf{p}} - e\mathbf{E}\cdot\frac{\partial \delta f}{\partial \mathbf{p}}
\end{align}
where we have used $\mathbf{v}\cdot\partial f_0/\partial \mathbf{x} = - \nabla \mu\cdot (\partial f_0/\partial \mathbf{p})$; this transformation makes it explicit that $\nabla \mu$ is a long-range force.

As seen from (\ref{eqn:delta_f}), in the presence of Coulomb interactions, the effective long-range force to $\mathcal{O}(\epsilon)$ is $-\nabla \bar{\mu} \equiv -\nabla \left(\mu - eV\right)$. Further, (\ref{eqn:fe_evol}) and (\ref{eqn:delta_f}) are the same at linear order. The last term in (\ref{eqn:delta_f}) which is not included in (\ref{eqn:fe_evol}) is only $\mathcal{O}(\epsilon^2)$ and is therefore ordered out.

The equivalence at linear order between (\ref{eqn:fe_evol}) and (\ref{eqn:delta_f}) allows for a convenient computational scheme in which we can solve the Boltzmann equation in a manner decoupled from the Poisson equation, while still taking into account the effect of long-range Coulomb interactions. We simply begin with an equilibrium $f_0$ set to a desired doping and setup transport by imposing a non-equilibrium $f$ at the location of the driving contacts. The price to pay however, is that we can only measure $\bar{\mu}$ from the simulation and not $V$ (or $\mu$).

A current-voltage experiment in fact measures $V$. To obtain this, we still need to solve the 3D Poisson equation with the appropriate boundary conditions and sourced by the 2D carrier density $n(\mathbf{x}) = \langle f \rangle$, where $f$ is computed from (\ref{eqn:fe_evol}) (or equivalently (\ref{eqn:delta_f})). Notably, this computation is a post-processing step and is not required for the evolution. As described in \S\ref{sec:IV}, we use an approximate solution to the 3D Poisson equation (\ref{eqn:LCA}) for the FET geometry that is widely used in experiments to obtain $V$.

\subsection{Drift vs Diffusion currents} \label{sec:drift_vs_diffusion}

We now proceed to check the relative contribution of each term in the total long-range force $-\nabla \bar{\mu}$. The term $-\nabla \mu$ gives rise to a diffusion current whereas $e\nabla V$ generates a drift current. Using (\ref{eqn:LCA}) to express $\nabla V$ in terms of $\nabla n$ and writing $\nabla \mu = (\partial \mu/\partial n) \nabla n$, we have,
\begin{align}
\nabla \bar{\mu} & = \nabla \mu - e \nabla V \\
                 & = \left(\frac{e^2}{C\sub{Q}} + \frac{e^2}{C}\right) \nabla n
\end{align}
where $C\sub{Q} = e^2 \partial n/\partial \mu$ is the quantum capacitance \cite{GrapheneCq}. We evaluate $C\sub{Q}$ for graphene using $\mu = \hbar v\sub{F} \sqrt{\pi n}$ at an equilibrium carrier density $n_0$ corresponding to the Fermi energy $\mu\sub{F}$. Using $C = \epsilon\sub{r}/(4\pi d)$ for a single gated device with a dielectric substrate $\epsilon\sub{r}$ and thickness $d$, the ratio of the drift to the diffusion term is,
\begin{align} \label{eqn:ratio}
8\cdot\alpha\sub{ee}\cdot (k\sub{F}d) \gg 1
\end{align}
where $\alpha\sub{ee} = e^2/(\epsilon\sub{r}\hbar v\sub{F}) \sim \mathcal{O}(1)$ is a dimensionless number called the graphene fine-structure constant and $k\sub{F} = \mu\sub{F}/(\hbar v\sub{F})$ is the Fermi wavenumber. With a typical $d\sim 100$ nm, we have $k\sub{F}d \gg 1$. From (\ref{eqn:ratio}), we see that the drift current dominates.

\subsection{Two different approximations to compute $V$}
With the diffusion current being negligible, a simple way to obtain voltages is to approximate $-e\nabla V \approx \nabla\bar{\mu}$, thus avoiding the Poisson equation altogether. Recall that $\bar{\mu}$ (written as $\mu$ in the main text) is obtained at every time step as a solution of the constraints (\ref{eqn:mass_cons}, \ref{eqn:energy_cons}, \ref{eqn:mom_cons}) that match $f$ to a local equilibrium $f_0$. The voltage thus obtained is indeed consistent with that obtained using the approximate solution to the Poisson equation (\ref{eqn:LCA}), since infact (\ref{eqn:LCA}) has been originally used in \S\ref{sec:drift_vs_diffusion} to arrive at the conclusion that the diffusion current is negligible.

\section{Incompressible Fluid Models} \label{sec:appendix_B}
Several calculations of current-voltage relations based on fluid models (e.g., \cite{LF2016, Torre2015, WhirlpoolsOrNot}) assume that the electron fluid is incompressible, and therefore $\delta n = 0$. However, in a kinetic calculation, density perturbations are inevitable since they appear at linear order, $\delta n = \langle \delta f \rangle \sim \mathcal{O}(\epsilon)$. Crucially, in this paper, these compressible fluctuations $\delta n$ are used to compute $V$ using an approximate solution (\ref{eqn:LCA}) to the 3D Poisson equation. Two questions arise: (1) How are voltages calculated when incompressibility is assumed?, and (2) Are they consistent with the approach to compute voltages followed in this paper which explicitly tracks compressible fluctuations?

Consider linearized fluid models described by the charge (\ref{eqn:charge_cty}) and momentum (\ref{eqn:momentum_cty}) conservation equations,
\begin{align}
\frac{\partial n}{\partial t} & = - n_0\nabla\cdot\mathbf{v}\sub{d} \label{eqn:charge_cty}\\
\frac{\partial\mathbf{v}\sub{d}}{\partial t} &= - \frac{1}{mn_0}\nabla P - \frac{e}{m}\mathbf{E} + \nu\nabla^2\mathbf{v}\sub{d} - \frac{\mathbf{v}\sub{d}}{\tau\sub{mr}} \label{eqn:momentum_cty}
\end{align}
where $n_0$ is the background charge density, $P$ is the pressure, $m$ is an effective mass, and $\nu$ is the kinematic viscosity of the electron fluid. The last term in (\ref{eqn:momentum_cty}) is a momentum loss term parametrized by a phenomenological timescale $\tau\sub{mr}$. The model requires a closure relation for the pressure $P$, which to a good approximation in the $\mu \gg kT$ limit is $P \simeq n\mu_F$. Equations (\ref{eqn:charge_cty}) and (\ref{eqn:momentum_cty}) are adopted from \cite{Torre2015, WhirlpoolsOrNot}, but with the time-derivatives and the pressure term included here.

The model thus has three unknowns, $(n, \mathbf{v}\sub{d}, V)$, and therefore requires three equations; the fluid equations (\ref{eqn:charge_cty}, \ref{eqn:momentum_cty}) along with the 3D Poisson equation (\ref{eqn:poisson}). However, the form of the force in (\ref{eqn:momentum_cty}) allows for the unknowns to be reduced by absorbing $P/(mn_0)$ and $-(e/m) V$ into a single scalar $\Phi = P/(mn_0) - (e/m)V$. The fluid system is then forced by the gradient of this effective scalar, analogous to $\bar{\mu}$ in the linearized kinetic equation (\ref{eqn:delta_f}). 

Rewriting (\ref{eqn:charge_cty}) and (\ref{eqn:momentum_cty}) in terms of $\Phi$ and taking the steady-state limit ($\partial/\partial t \rightarrow 0$),
\begin{align}
\nabla\cdot\mathbf{v}\sub{d} & = 0 \label{eqn:charge_cty_steady_state}\\
- \nabla \Phi + \nu \nabla^2\mathbf{v}\sub{d} & = \frac{\mathbf{v}\sub{d}}{\tau\sub{mr}}\label{eqn:momentum_cty_steady_state}
\end{align}
Note that the steady-state limit removes the dependence of $n$ in (\ref{eqn:charge_cty}). There are now only two variables $(\mathbf{v}\sub{d}, \Phi)$, for which just the equations (\ref{eqn:charge_cty_steady_state}, \ref{eqn:momentum_cty_steady_state}) are sufficient. This is similar to the kinetic scheme discussed in appendix \S\ref{sec:appendix_A} where the 3D Poisson equation is not necessary to compute $f$ to linear order.

However, just as described in \S\ref{sec:appendix_A}, the reduced system only yields $\Phi$ and not $V$. At this point, one can invoke incompressibility ($n = n_0 \equiv$ constant) to set $\nabla P = 0$. This immediately reduces the unknowns to $(\mathbf{v}\sub{d}, V)$, thus fully determining the current-voltage characteristics. It is then not necessary to solve the 3D Poisson equation (\ref{eqn:poisson}), which only determines the background carrier density $n_0$.

The incompressible approximation is equivalent to stating that the diffusion current that arises due to compressible fluctuations is zero. This is consistent with \S\ref{sec:drift_vs_diffusion} where it is seen that although both diffusion and drift currents appear at linear order in a kinetic calculation, the ratio of the prefactors (\ref{eqn:ratio}) are such that diffusion is negligible. Further, this ratio has been computed using (\ref{eqn:LCA}), which we use to obtain voltages in the main text. The voltages thus obtained are therefore consistent with those obtained by fluid models in the incompressible limit (whenever the fluid models are applicable: $l\sub{mc},l\sub{mr} \ll L$).

\begin{acknowledgments}
We thank Shyam Sankaran for design and performance optimizations in {\tt bolt} that allowed for an efficient parameter exploration.
We thank Krishnendu Chatterjee and Joshua Matthew for insightful discussions.
MC, GK and DS thank {\it Quazar Technologies} for funding this research effort.
RS acknowledges start-up funding from the Department of Materials Science and Engineering and
computational resources at the Center for Computational Innovations at Rensselaer Polytechnic Institute.
\end{acknowledgments}


\begin{thebibliography}{5}

\bibitem[Ashcroft \& Mermin(1976)]{AshMer} N. Ashcroft, \& D. Mermin, \emph{Solid State Physics}, Brooks/Cole, 1976

\bibitem[Mayorov et al.(2011)]{GrapheneMicronBallistic} A. S. Mayorov, R. V. Gorbachev, S. V. Morozov, {\it et al.}, {\it Nano Letters} 2011 11 (6), 2396-2399

\bibitem[de Jong \& Molenkamp(1995)]{deJong} M.~J.~M. de Jong, \& L.~W. Molenkamp, {\it Phys. Rev. B} {\bf 51}, 13389 (1995)

\bibitem[Molenkamp \& de Jong(1994)]{Molenkamp} L.~W. Molenkamp, \& M.~J.~M. de Jong, {\it Phys. Rev. B} {\bf 49}, 5038 (1994)

\bibitem[Braem et al.(2018)]{ScanningGateGaAs}B.~A. Braem, F.~M.~D. Pellegrino, A. Principi, {\it et al.}, {\it Phys. Rev. B} {\bf 98}, 241304(R) (2018)

\bibitem[Gusev et al.(2018)]{ViscousFlow2DGas} G.~M. Gusev, A.~D. Levin, E.~V. Levinson, and A.~K. Bakarov, AIP Advances, 8, 025318 (2018)

\bibitem[Moll et al.(2016)]{PdCoO2} P.~J.~W. Moll, P. Kushwaha, N. Nandi, B. Schmidt, and A. P. Mackenzie, {\it Science} {\bf 351}, 1061-1064 (2016)

\bibitem[Gooth et al.(2018)]{WP2} J. Gooth, F. Menges, N. Kumar, {\it et al.}, {\it Nature Comm.} {\bf 9}, 4093 (2018)

\bibitem[Bandurin et al.(2016)]{NegativeLocal} D.~A. Bandurin, I. Torre, R. Krishna Kumar, {\it et al.}, {\it Science} {\bf 351}, 1055-1058 (2016)

\bibitem[Krishna Kumar et al.(2017)]{SuperBallisticExpt} R. Krishna Kumar, D.~A. Bandurin, F.~M.~D. Pellegrino, {\it et al.}, {\it Nature Physics} {\bf 13}, 1182 (2017)

\bibitem[Bandurin et al.(2018)]{FluidityOnsetExpt} D.~A. Bandurin, Shytov, A., Levitov, L., {\it et al.}, {\it Nature Comm.} {\bf 9}, 4533 (2018)

\bibitem[Berdyugin et al.(2019)]{GrapheneMHD} A.~I. Berdyugin et al., {\it Science} 10.1126/science.aau0685 (2019)

\bibitem[Gurzhi(1968)]{Gurzhi} R.~N. Gurzhi, Soviet Physics Uspekhi, 11, 255 (1968)

\bibitem[Guo et al.(2017)]{SuperBallisticTheory} H. Guo, E. Ilseven, G. Falkovich, \& L. Levitov, Proceedings of the National Academy of Science, 114, 3068 (2017)

\bibitem[Guo et al.(2017)]{StokesParadox1} H. Guo, E. Ilseven, G. Falkovich, \& L. Levitov, arxiv:1612.09239

\bibitem[Lucas(2017)]{StokesParadox2} A. Lucas, {\it Phys. Rev. B} {\bf 95}, 115425 (2017)

\bibitem[Shytov et al.(2018)]{FluidityOnsetTheory} Shytov, A., Kong, J. F., Falkovich, G., and Levitov, L., {\it Phys. Rev. Lett.} {\bf 121}, 176805 (2018)

\bibitem[Alekseev(2016)]{NegativeMagnetoresistance} P.~S. Alekseev, {\it Phys. Rev. Lett.} {\bf 117}, 166601 (2016)

\bibitem[Principi \& Vignale(2015)]{ViolationWF} A. Principi, \& G. Vignale, {\it Phys. Rev. Lett.} {\bf 115}, 056603 (2015)

\bibitem[Scaffidi et al.(2017)]{HydroHallViscosity1} T. Scaffidi, N. Nandi, B. Schmidt, A.~P. Mackenzie, \& J.~E. Moore, {\it Phys. Rev. Lett.} {\bf 118}, 226601 (2017)

\bibitem[Delacr\`etaz \& Gromov(2017)]{HydroHallViscosity2} L.~V. Delacr\`etaz, \& A. Gromov, {\it Phys. Rev. Lett.} {\bf 119}, 226602 (2017)

\bibitem[Pellegrino et al.(2017)]{HydroHallViscosity3} F.~M.~D. Pellegrino, I. Torre, \& M. Polini, {\it Phys. Rev. B} {\bf 96}, 195401 (2017)


\bibitem[Chow et al.(1996)]{PhononEmission} E. Chow, H. P. Wei, S. M. Girvin, \& M. Shayegan, {\it Phys. Rev. Lett.} {\bf 77}, 1143 (1996)

\bibitem[Govorov \& Heremans(2004)]{FermiJets} A.~O. Govorov, \& J.~J. Heremans, {\it Phys. Rev. Lett.} {\bf 92}, 026803 (2004)


\bibitem[Tomadin et al.(2013)]{Tomadin2013} A. Tomadin, \& M. Polini, {\it Phys. Rev. B} {\bf 88}, 205426 (2013)

\bibitem[Tomadin et al.(2014)]{CorbinoDisc} A. Tomadin, G. Vignale, \& M. Polini, {\it Phys. Rev. Lett.} {\bf 113}, 235901 (2014)

\bibitem[Lucas(2016)]{SoundModes} A. Lucas, {\it Phys. Rev. B} {\bf 93}, 245153 (2016)

\bibitem[Semenyakin et al.(2018)]{FalkovichAC} M. Semenyakin \& G. Falkovich, {\it Phys. Rev. B} {\bf 97}, 085127 (2018)

\bibitem[Levitov \& Falkovich(2016)]{LF2016} L. Levitov, \& G. Falkovich, {\it Nature Physics} {\bf 12}, 672 (2016)

\bibitem[Torre et al.(2015)]{Torre2015} I. Torre, A. Tomadin, A.~K. Geim, \& M. Polini, {\it Phys. Rev. B} {\bf 92}, 165433 (2015)

\bibitem[Pellegrino et al.(2016)]{WhirlpoolsOrNot} F.~M.~D. Pellegrino, I. Torre, A.~K. Geim, \& M. Polini, {\it Phys. Rev. B} {\bf 94}, 155414 (2016)

\bibitem[Falkovich \& Levitov(2017)]{LF2017} G. Falkovich, \& L. Levitov, {\it Phys. Rev. Lett.} {\bf 119}, 066601 (2017)

\bibitem[Dyakonov \& Shur(1993)]{DyaShu} M. Dyakonov and M. Shur, {\it Phys. Rev. Lett.} {\bf 71}, 2465 (1993)

\bibitem[Svintsov et al.(2012)]{HydroElectronHolePlasma} D. Svintsov, V. Vyurkov, S. Yurchenko, T. Otsuji and V. Ryzhii, {\it Journal of Applied Physics} {\bf 111}, 083715 (2012)

\bibitem[Svintsov(2018)]{HydroToBallisticCrossoverDirac} D. Svintsov, {\it Phys. Rev. B} {\bf 97}, 121405(R) (2018)

\bibitem[Gabbana et al.(2018)]{PreTurbulenceProspects} A. Gabbana, M. Polini, S. Succi, R. Tripiccione, and F.~M.~D. Pellegrino, {\it Phys. Rev. Lett.} {\bf 121}, 236602 (2018)

\bibitem[Chandra et al.(in prep)]{bolt} M. Chandra, S. Sankaran, P. Yalamanchili (in prep), \url{www.quazartech.com/bolt}

\bibitem[Kraichnan(1975)]{Enstrophy} R.~H. Kraichnan, \ 1975, Journal of Fluid Mechanics, 67, 155 

\bibitem[Lozovik et al.(2015)]{GrapheneCq} Yu.~E. Lozovik, A.~A. Sokolik, and A.~D. Zabolotskiy, {\it Phys. Rev. B} {\bf 91}, 075416 (2015)

\bibitem[Andreev et al.(2011)]{HydroStronglyCorrelated} A.~V. Andreev, S.~A. Kivelson, and B. Spivak, {\it Phys. Rev. Lett.} {\bf 106}, 256804 (2011)

\bibitem[Crossno et al.(2016)]{DiracFluid} J. Crossno, J.~K. Shi, K. Wang, {\it et al.}, {\it Science} {\bf 351}, 1058-1061 (2016)

\bibitem[Gallagher et al.(2019)]{GrapheneQuantumCriticalConductivity} P. Gallagher et al., {\it Science} 10.1126/science.aat8687 (2019)

\bibitem[Lucas et al.(2016)]{DiracFluidTheory} A. Lucas, J. Crossno, K.~C. Fong, P. Kim, \& S. Sachdev, {\it Phys. Rev. B} {\bf 93}, 075426 (2016)

\bibitem[M{\"u}ller et al.(2009)]{GraphenePerfectFluid} M. M{\"u}ller, J. Schmalian, \& L. Fritz, {\it Phys. Rev. Lett.} {\bf 103}, 025301 (2009)

\end{thebibliography}
\end{document}